\documentclass[aps,twocolumn]{revtex4}
\usepackage{graphicx}
\usepackage{epsfig}
\usepackage{dcolumn}
\usepackage{amsmath}
\def\Journal#1#2#3#4{{#1} {\bf#2}, #3 (#4)}

\def\NPA{{\rm Nucl. Phys.} A}
\def\NPB{{\rm Nucl. Phys.} B}
\def\PLB{{\rm Phys. Lett.}  B}
\def\PRL{\rm Phys. Rev. Lett.}
\def\PRD{{\rm Phys. Rev.} D}
\def\PRC{{\rm Phys. Rev.} C}

\def\RMP{\rm Rev. Mod. Phys.}

\def\ep{\epsilon}
\def\lam{\lambda}

\def\la{\langle}
\def\ra{\rangle}

\def\al{\alpha}

\def\be{\begin{equation}}
\def\ee{\end{equation}}
\def\bea{\begin{eqnarray}}
\def\eea{\end{eqnarray}}
\begin{document}
\title{Light-front quark model analysis of the exclusive rare
$B_c\to D_{(s)}(\ell^+\ell^-,\nu_{\ell}\bar{\nu}_{\ell})$  decays}
\author{ Ho-Meoyng Choi\\
Department of Physics, Teachers College, Kyungpook National
University, Daegu, Korea 702-701}
\begin{abstract}
We investigate the exclusive rare $B_c\to D_{(s)}\nu_{\ell}\bar{\nu_{\ell}}$ and
$B\to D_{(s)}\ell^+\ell^-$ ($\ell=e,\mu,\tau$) decays within the standard model
and the light-front quark model constrained by
the variational principle for the QCD motivated effective Hamiltonian. The form factors
$f_{\pm}(q^2)$ and $f_T(q^2)$ are obtained from the analytic continuation method in
the $q^+=0$ frame. While the form factors $f_+(q^2)$ and $f_T(q^2)$ are free from the zero-mode,
the form factor $f_-(q^2)$ is not free from the zero-mode in the $q^+=0$
frame. We discuss the covariance(i.e. frame-independence) of our model calculation and
quantify the zero-mode contributions to $f_-(q^2)$ for $B_c\to D_{(s)}$ decays. The
branching ratios and the longitudinal lepton polarization asymmetries are calculated with
and without the long-distance contributions. Our numerical results for the non-resonant branching ratios for
$B_c\to D(D_s)\sum\nu_{\ell}\bar{\nu_{\ell}}$ and
$B_c\to D(D_s)\ell^+\ell^-$
are in the order of $10^{-8}\;(10^{-7})$ and $10^{-9}\;(10^{-8})$, respectively.
The averaged values of the  lepton polarization asymmetries
obtained from the linear (harmonic oscillator) potential parameters
are found to be $-0.99\;(-0.99)$ for $B_c\to D\mu^+\mu^-$  and $-0.16\;(-0.15)$ for
$B_c\to D\tau^+\tau^-$, and $-0.98\;(-0.98)$ for $B_c\to D_s\mu^+\mu^-$ and $-0.14\;(-0.12)$ for
$B_c\to D_s\tau^+\tau^-$, respectively.
\end{abstract}


\maketitle
\section{Introduction }
The LHCb (Large Hadron Collider beauty) experiment dedicated to heavy flavor physics at LHC
make precision tests of the standard model (SM) and beyond the SM ever more promising.
Its primary goal is to look for indirect evidence of new physics in CP violation and rare
decays of beauty and charm hadrons.
Especially, a stringent test on the unitarity of
Cabibbo-Kobayashi-Maskawa (CKM) mixing matrix in the SM
will be made by this facility.
With the upcoming chances that a numerous number of $B_c$ mesons will be produced
at LHC, one might explore the exclusive rare $B_c$ decays to
$(D,D_s)\ell^+\ell^-$ and $(D,D_s)\nu_{\ell}\bar{\nu}_{\ell}$ ($\ell=e,\mu,\tau$)
induced by the flavor-changing neutral current $b\to(d,s)$ transitions. 
Since in the SM the rare $B_c$ decays are
forbidden at tree level and occur at the lowest order only through one-loop
diagrams~\cite{GWS,BM,Misiak,TI,AMM,KMS,AKS}, they are well suited to test the SM and search
for physics beyond the SM.  In such exclusive rare decays, any reliable extraction of the
perturbative effects encoded in the Wilson coefficients of the effective Hamiltonian
requires an accurate separation of the nonperturbative contributions, which are encoded
in the hadronic form factors. This part of the calculation is model dependent since it
involves nonperturbative QCD. Therefore, a reliable estimate of the hadronic form factors
for the exclusive rare $B_c$ decays is very important to make correct predictions within
and beyond the SM.

There are some theoretical approaches to the calculations of the exclusive rare
$B_c\to (D,D_s)\ell^+\ell^-$
and $B_c\to (D,D_s)\nu_{\ell}\bar{\nu}_{\ell}$ decay modes.
Although we may not be
able to list them all, we may note here the following works: the
relativistic constituent quark model~\cite{Faessler}, the light-front(LF) and constituent
quark model (CQM)~\cite{Geng}, and three point QCD sum rules~\cite{Azizi}. The rare
$B_c\to D_s\ell^+\ell^-$ decay beyond the SM has also been studied in~\cite{Yilmaz}.
Perhaps, one of the most well-suited formulations for the analysis of exclusive processes
involving hadrons may be provided in the framework of light-front quantization~\cite{BPP}.

The purpose of this paper is to extend our our light-front quark
model (LFQM)~\cite{CJ1,CJ_PLB1,JC_E,CJK,ChoiRD,CJBc,CJNRD}
based on the QCD-motivated effective LF Hamiltonian to calculate the hadronic form factors,
decay rates and the longitudinal lepton polarization asymmetries (LPAs)
for the exclusive rare $B_c\to (D,D_s)\ell^+\ell^-$
and $B_c\to (D,D_s)\nu_{\ell}\bar{\nu}_{\ell}$ decays within the SM.
The LPA, as another parity-violating
observable, is an important asymmetry~\cite{Hew} and could be
measured by the LHCb experiment.
In particular, the $\tau$ channel would be more accessible
experimentally than $e$- or $\mu$-channels since the LPAs
in the SM are known to be proportional to the lepton
mass.

In our previous LFQM analysis~\cite{CJBc,CJNRD}, we have analyzed the exclusive semileptonic
$B_c\to (D,\eta_c,B,B_s)\ell\nu_\ell$ decays~\cite{CJBc} and the nonleptonic two-body
decays of $B_c$ mesons such as $B_c\to (D_{(s)},\eta_c,B_{(s)})(P,V)$ decays~\cite{CJNRD}
(here $P$ and $V$ denote pseudoscalar and vector mesons, respectively).
Our LFQM~\cite{CJ1,CJ_PLB1,JC_E,CJK,ChoiRD,CJBc,CJNRD} analysis compared to the other LFQM
has several salient features: (i) We have implemented the
variational principle to the QCD motivated effective LF
Hamiltonian to enable us to analyze the meson mass spectra and to
find optimized model parameters~\cite{CJ1,CJ_PLB1}. (ii)
The weak form factors $f_{\pm}(q^2)$ for the semileptonic decays
between two pseudoscalar mesons are obtained in the Drell-Yan-West ($q^+=q^0+q^3=0$) frame~\cite{DYW}
(i.e., $q^2=-{\bf q}^2_\perp<0$) and then analytically continued to the timelike
region by changing ${\bf q}^2_\perp$ to $-q^2$ in the form factor.
The covariance (i.e., frame independence) of our model has been checked~\cite{CJBc} by performing the
LF calculation in the $q^+=0$ frame in parallel with the manifestly
covariant calculation using the exactly solvable covariant fermion
field theory model in $(3+1)$ dimensions. We also found the zero-mode~\cite{Zero} contribution
to the form factor $f_-(q^2)$ and identified~\cite{CJBc} the zero-mode operator that
is convoluted with the initial and final state LF wave functions.

Specifically, in the present analysis of exclusive rare $B_c\to (D,D_s)\ell^+\ell^-$
and $B_c\to (D,D_s)\nu_{\ell}\bar{\nu}_{\ell}$ decays, three independent hadronic form
factors, i.e. $f_+(q^2)$, $f_-(q^2)$ from the vector-axial vector current,
and $f_T(q^2)$ from the tensor current, are needed.
While the two form factors $f_+$ and $f_T$
can be obtained only from the valence contributions in the $q^+=0$ frame
without encountering the zero-mode complication,
the form factor $f_-(q^2)$ receives the higher Fock state contribution (i.e.,
the zero mode in the $q^+=0$ frame or the nonvalence contribution
in the $q^+> 0$ frame) within the framework of LF quantization. Thus, it
is necessary to include either the zero-mode contribution (if
working in the $q^+=0$ frame) or the nonvalence contribution (if
working in the $q^+> 0$ frame) to obtain the form factor $f_-(q^2)$.
In this work, we shall use the form factors $f_+(q^2)$ and $f_-(q^2)$ for the exclusive
semileptonic $B_c\to (D, D_s)$ decays obtained in~\cite{CJBc} and the form factor
$f_T(q^2)$ obtained in~\cite{CJK} for the analysis of $B\to K\ell^+\ell^-$ decay.
Especially, the Lorentz covariance of our tensor form factor
$f_T(q^2)$ is discussed in this work.
The present investigation further constrains the phenomenological parameters and extends
the applicability of our LFQM~\cite{CJ1,CJ_PLB1} to the wider range of hadronic phenomena.

The paper is organized as follows. In Sec. II, the SM operator basis,
describing the $b\to (d,s)\ell^+\ell^-$ and $b\to (d,s)\nu_{\ell}\bar{\nu}_\ell$
transitions, is briefly presented.  
In Sec. III, we  briefly
describe the formulation of our LFQM and the procedure of fixing
the model parameters using the variational principle for the QCD
motivated effective Hamiltonian. We present the
LF covariant forms of the form factors $f_{\pm}(q^2)$ and
$f_T(q^2)$ obtained in the $q^+=0$ frame.
In Sec. IV, our numerical results, i.e. the form factors, decay rates,
and the LPAs for the
rare $B_c\to (D,D_s)\ell^+\ell^-$
and  $B_c\to (D,D_s)\nu_{\ell}\bar{\nu}_\ell$   decays
are presented. Summary and discussion of our main results follow in Sec. V.
In the Appendix, we explicitly show the covariance of $f_T(q^2)$
by performing the LF calculation in parallel with the manifestly covariant one 
using the exactly solvable covariant fermion
field theory model in $(3+1)$ dimensions.

\section{Effective Hamiltonian}
\begin{figure}
\includegraphics[width=3.0in,height=1.2in]{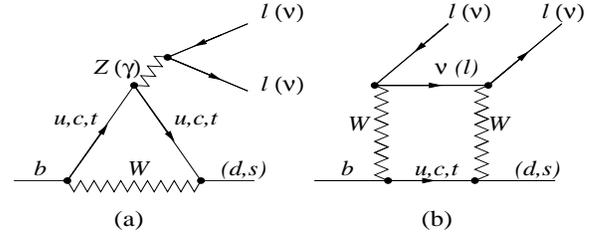}
\caption{ Loop diagrams for $B_c\to D_q(\ell^+\ell^-,\nu_\ell\bar{\nu}_\ell)(q=d,s)$
transitions. \label{fig1_BcDs} }
\end{figure}

In the SM, the exclusive rare $B_c\to D_q(\ell^+\ell^-, \nu_\ell\bar{\nu}_\ell)$
decays are at the quark level described by the loop
$b\to q(\ell^+\ell^-, \nu_\ell\bar{\nu}_\ell)(q=d,s)$ transitions, and receive
contributions from the $Z(\gamma)$-penguin and $W$-box diagrams as shown in Fig.~\ref{fig1_BcDs}.

The effective Hamiltonian responsible for the $b\to q\ell^+\ell^-(q=d,s)$ decay processes
can be represented in terms of the Wilson coefficients, $C_7^{\rm eff},C_9^{\rm eff}$
and $C_{10}$ as~\cite{BM}
\bea\label{Hll}
{\cal H}^{\ell^+\ell^-}_{\rm eff}&=&
\frac{G_F\al_{\rm em}}{2\sqrt{2}\pi}V_{tb}V^*_{tq}
\biggl[C^{\rm eff}_9\bar{q}\gamma_\mu (1-\gamma_5) b\bar{\ell}\gamma^\mu\ell
\nonumber\\
&&+ C_{10}\bar{q}\gamma_\mu (1-\gamma_5) b\bar{\ell}\gamma^\mu\gamma_5\ell
\nonumber\\
&&-C_7^{\rm eff}\frac{2m_b}{q^2}\bar{q}i\sigma_{\mu\nu}q^\nu(1+\gamma_5) b
\bar{\ell}\gamma^\mu\ell
\biggr],
\eea
where $G_F$ is the Fermi constant, $\alpha_{\rm em}$ is the fine structure
constant, and $V_{ij}$ are the CKM matrix elements. The relevant Wilson
coefficients $C_i$ can be found in Ref.~\cite{BM}.
The effective Hamiltonian responsible for the $b\to q\nu_{\ell}\bar{\nu}_{\ell}(q=d,s)$
decay processes is given by~\cite{MN}
\bea\label{Hnn}
{\cal H}^{\nu_{\ell}\bar{\nu}_{\ell}}_{\rm eff}&=&
\frac{G_F\al_{\rm em}}{2\sqrt{2}\pi}V_{tb}V^*_{tq}\frac{X(x_t)}{\sin^2\theta_W}
\bar{q}\gamma_\mu (1-\gamma_5) b
\nonumber\\
&\times&\bar{\nu}_{\ell}\gamma^\mu(1-\gamma_5)\nu_{\ell},
\eea
where $x_t=(m_t/M_W)^2$ and $X(x_t)$ is an Inami-Lim function~\cite{TI}, which
is given by
\be\label{ILF}
X(x) = \frac{x}{8}\biggl( \frac{2+x}{x-1} + \frac{3x-6}{(x-1)^2}\ln x\biggr).
\ee
The long distance (LD) contribution to the exclusive $B_c\to D_q (q=d,s)$ decays is
contained in the meson matrix elements of the bilinear quark currents
appearing in ${\cal H}^{\ell^+\ell^-}_{\rm eff}$ and
${\cal H}^{\nu_{\ell}\bar{\nu_{\ell}}}_{\rm eff}$.
In the matrix elements
of the hadronic currents for $B_c\to D_{q}$ transitions, the parts containing
$\gamma_5$ do not contribute. Considering Lorentz and parity invariances,
these matrix elements can be parametrized in terms of hadronic form factors
as follows:
\be\label{Jmu}
J^\mu\equiv\la D_q|\bar{q}\gamma^{\mu} b|B_c\ra=
f_{+}(q^{2})P^\mu + f_{-}(q^{2})q^\mu,
\ee
and
\bea\label{JTmu}
J^\mu_T&\equiv&\la D_q|\bar{q}i\sigma^{\mu\nu}q_\nu b|B_c\ra
\nonumber\\
&=& \frac{f_T(q^2)}{M_{B_c}+M_{D_q}}[q^2 P^\mu - (M^2_{B_c}-M^2_{D_q})q^\mu],
\eea
where $P=P_{B_c}+P_{D_q}$ and $q=P_{B_c}-P_{D_q}$ is the four-momentum
transfer to the lepton pair and $4m^{2}_{\ell}\leq q^{2}\leq(M_{B_c}-M_{D_q})^2$.
We use the convention $\sigma^{\mu\nu}=(i/2)[\gamma^\mu,\gamma^\nu]$ for
the antisymmetric tensor.
Sometimes it is useful to express Eq.~(\ref{Jmu}) in terms
of $f_+(q^2)$ and $f_0(q^2)$, which are related to the exchange
of $1^-$ and $0^+$, respectively, and satisfy the following relations:
\be\label{F0}
f_+(0)=f_0(0),\;
f_0(q^2)=f_+(q^2) + \frac{q^2}{M^2_{B_c}-M^2_{D_q}}f_-(q^2).
\ee
With the help of the effective Hamiltonian in Eq.~(\ref{Hll}) and
Eqs.~(\ref{Jmu}) and~(\ref{JTmu}),  the transition amplitude
${\cal M}= \la D_q\ell^+\ell^-|{\cal H}_{\rm eff}|B_c\ra$
for the $B_c\to D_q\ell^+\ell^-$ decay can be written as
\bea\label{TranA}
{\cal M}
&=&\frac{G_F\al_{\rm em}}{2\sqrt{2}\pi}V_{tb}V^*_{tq}
\biggl\{\biggl[C^{\rm eff}_9 J_\mu - \frac{2m_b}{q^2}C^{\rm eff}_7 J^T_\mu\biggr]
\bar{\ell}\gamma^\mu\ell \nonumber\\
&&\hspace{2.5cm}
+ C_{10}J_\mu \bar{\ell}\gamma^\mu\gamma_5\ell \biggr\}.
\eea
The differential decay rate for the exclusive
rare $B_c\to D_q\ell^+\ell^-$ with nonzero lepton mass
is given by~\cite{GK,MN}
\bea\label{DDR}
\frac{d\Gamma}{ds}
&=&\frac{M^5_{B_c}G^2_F}{3\cdot2^9\pi^5}\alpha^2_{\rm em}
|V_{tb}V^*_{tq}|^2\phi_H^{1/2}
\biggl(1-\frac{4t}{s}\biggr)^{1/2}
\nonumber\\
&&\times
\biggl[\phi_H\biggl(1+\frac{2t}{s}\biggr)
{\cal F}_{1} + 12t{\cal F}_{2} \biggr],
\eea
where
\bea\label{DDR2}
{\cal F}_{1} &=&
\biggl|C^{\rm eff}_9 f_+
- \frac{2\hat{m_b}C^{\rm eff}_7}{1+\sqrt{r}}f_T \biggr|^2
+ |C_{10}f_+|^2,\nonumber\\
{\cal F}_{2} &=&
|C_{10}|^2 \biggl[(1+r -\frac{s}{2})|f_+|^2
+ (1-r)f_+f_-
+\frac{s}{2}|f_-|^2\biggr],
\nonumber\\
\phi_H  &=& (s-1-r)^2-4r,
\nonumber\\
\eea
with $s=q^2/M^2_{B_c}$, $t=m^2_\ell/M^2_{B_c}$,
$\hat{m_b}=m_b/M_{B_c}$ and $r=M^2_{D_q}/M^2_{B_c}$.
The differential decay rate in Eq.~(\ref{DDR}) may be written in
terms of ($f_+,f_0,f_T$) instead of ($f_+,f_-,f_T$) as discussed
in~\cite{CJK}. Note also from Eqs.~(\ref{DDR}) and~(\ref{DDR2}) that
the form factor $f_-(q^2)$ contributes
only in the nonzero lepton ($m_\ell\neq 0$) mass limit.
Dividing Eq.~(\ref{DDR}) by the total width
of the $B_c$ meson,
one can obtain the differential branching ratio
$d{\rm BR}(B_c\to D_q\ell^+\ell^-)/ds
=(d\Gamma(B_c\to D_q\ell^+\ell^-)/\Gamma_{\rm tot})/ds$.

The differential decay rate for $B_c\to D_q\nu_{\ell}\bar{\nu}_{\ell}$ can be easily
obtained from the corresponding formula Eq.~(\ref{DDR})
for $B_c\to D_q\ell^+\ell^-$ by the replacement
\be\label{Dnu}
\hat{m}_\ell\to0,\; C_7^{\rm eff}\to 0,\;
C^{\rm eff}_9\to\frac{X(x_t)}{\sin^2\theta_W},\;
C_{10}\to-\frac{X(x_t)}{\sin^2\theta_W},
\ee
where $\theta_W$ is the Weinberg angle.
As another interesting observable, the LPA, is defined as
\be\label{LPA}
P_L(s)=\frac{d\Gamma_{h=-1}/ds-d\Gamma_{h=1}/ds}
{d\Gamma_{h=-1}/ds +d\Gamma_{h=1}/ds},
\ee
where $h=+1\;(-1)$ denotes right (left) handed $\ell^-$ in the final state.
From Eq.~(\ref{DDR}), one obtains for $B_c\to D_q\ell^+\ell^-$
\be\label{LPA_Bc}
P_L(s)=\frac{
2\biggl(1-4\frac{t}{s}\biggr)^{1/2}
\phi_H C_{10}f_{+}
\biggl[f_+ {\rm Re}C^{\rm eff}_9 -
\frac{2\hat{m_b}C^{\rm eff}_7}{1+\sqrt{r}}f_T\biggr] }
{ \biggl[\phi_H\biggl(1+2\frac{t}{s}\biggr){\cal F}_{1}
+ 12t{\cal F}_{2} \biggr] }.
\ee
Because of the experimental difficulties of studying the polarizations
of each lepton depending on $s$ and the Wilson coefficients, it would be
better to eliminate the dependence of the LPA on $s$, by considering the
averaged form over the entire kinematical region.
The averaged LPA is defined by
\bea\label{LPA_av}
\la P_L\ra = \frac{\int^{(1-\sqrt{r})^2}_{4t} P_L\frac{dBR}{ds}ds}
{\int^{(1-\sqrt{r})^2}_{4t} \frac{dBR}{ds}ds}.
\eea

\section{Form factors in Light-front quark model}
The key idea in our LFQM~\cite{CJ1,CJ_PLB1} for mesons is to treat the
radial wave function as a trial function for the variational
principle to the QCD-motivated effective Hamiltonian saturating
the Fock state expansion by the constituent quark and antiquark.
The QCD-motivated Hamiltonian for a description of the ground
state meson mass spectra is given by
\bea\label{Ham}
H_{q\bar{q}}|\Psi^{JJ_z}_{nlm}\ra&=&\biggl[ \sqrt{m^2_q+{\vec
k}^2}+\sqrt{m^2_{\bar{q}}+{\vec k}^2}+V_{q\bar{q}}\biggr]
|\Psi^{JJ_z}_{nlm}\ra,
\nonumber\\
&=&[H_0 + V_{q\bar{q}}]|\Psi^{JJ_z}_{nlm}\ra
=M_{q\bar{q}}|\Psi^{JJ_z}_{nlm}\ra, \eea where ${\vec k}=({\bf
k}_\perp, k_z)$ is the three-momentum of the constituent quark,
$M_{q\bar{q}}$ is the mass of the meson, and
$|\Psi^{JJ_z}_{nlm}\ra$ is the meson wave function.
We use two interaction potentials $V_{q\bar{q}}$; (i) Coulomb
plus harmonic oscillator (HO) and (ii) Coulomb plus linear confining
potentials. The hyperfine interaction essential to
distinguish pseudoscalar and vector mesons
is also included; viz.,
\be\label{pot} V_{q\bar{q}}=V_0 + V_{\rm hyp} = a + {\cal V}_{\rm
conf}-\frac{4\al_s}{3r} +\frac{2}{3}\frac{{\bf S}_q\cdot{\bf
S}_{\bar{q}}}{m_qm_{\bar{q}}} \nabla^2V_{\rm coul},
\ee
where
${\cal V}_{\rm conf}=br\;(r^2)$ for the linear (HO) potential and
$\la{\bf S}_q\cdot{\bf S}_{\bar{q}}\ra=1/4\;(-3/4)$ for the vector
(pseudoscalar) meson. Using this Hamiltonian, we analyze the meson mass spectra and
various wave-function-related observables, such as decay
constants, electromagnetic form factors of mesons in a spacelike
region, and the weak form factors for the exclusive semileptonic
and rare decays of pseudoscalar mesons in the timelike
region~\cite{CJ1,CJ_PLB1,JC_E,CJK,ChoiRD,CJBc}.

The momentum-space LF wave function of the ground state
pseudoscalar mesons is given by
\be\label{w.f}
\Psi^{00}_{100}(x_i,{\bf k}_{i\perp},\lam_i) ={\cal
R}^{00}_{\lam_1\lam_2}(x_i,{\bf k}_{i\perp}) \phi(x_i,{\bf
k}_{i\perp}),
\ee
where $\phi(x_i,{\bf k}_{i\perp})$ is the radial
wave function and ${\cal R}^{00}_{\lam_1\lam_2}$ is the
spin-orbit wave function. The model wave function in Eq.~(\ref{w.f}) is
represented by the Lorentz-invariant internal variables, $x_i=p^+_i/P^+$,
${\bf k}_{i\perp}={\bf p}_{i\perp}-x_i{\bf P}_\perp$ and $\lam_i$,
where $P^\mu=(P^+,P^-,{\bf P}_\perp) =(P^0+P^3,(M^2+{\bf
P}^2_\perp)/P^+,{\bf P}_\perp)$ is the momentum of the meson $M$,
and $p^\mu_i$ and $\lam_i$ are the momenta and the helicities of
constituent quarks, respectively.
The covariant forms of the spin-orbit wave function
for pseudoscalar  mesons is given by
\bea\label{R00_A}
{\cal R}_{\lam_1\lam_2}^{00}
&=&\frac{-\bar{u}_{\lam_1}(p_1)\gamma_5 v_{\lam_2}(p_2)}
{\sqrt{2}\tilde{M_0}},
\eea
where $\tilde{M_0}=\sqrt{M^2_0-(m_1-m_2)^2}$ and
$M^2_0=\sum_{i=1}^2({\bf k}^2_{i\perp}+m^2_i)/x_i$ is
the boost invariant meson mass square obtained from the free
energies of the constituents in mesons.
For the radial wave function $\phi$, we use the
Gaussian wave function:
\be\label{rad}
 \phi(x_i,{\bf
k}_{i\perp})=\frac{4\pi^{3/4}}{\beta^{3/2}} \sqrt{\frac{\partial
k_z}{\partial x}} {\rm exp}(-{\vec k}^2/2\beta^2),
 \ee
 where $\beta$ is the variational parameter and $\sqrt{\partial k_z/\partial x}$
 is the Jacobian of the variable transformation
$\{x,{\bf k}_\perp\}\to {\vec k}=({\bf k}_\perp, k_z)$. 

We apply our
variational principle to the QCD-motivated effective Hamiltonian
first to evaluate the expectation value of the central Hamiltonian
$H_0+V_0$, i.e., $\la\phi|(H_0+V_0)|\phi\ra$ with a trial
function $\phi(x_i,{\bf k}_{i\perp})$ that depends on the
variational parameter $\beta$. Once the model
parameters are fixed by minimizing the expectation value
$\la\phi|(H_0+V_0)|\phi\ra$, then the mass eigenvalue of each meson is
obtained as $M_{q\bar{q}}=\la\phi|(H_0+V_{q\bar{q}})|\phi\ra$.
A more detailed procedure
for determining the model parameters of light- and heavy-quark
sectors can be found in our previous works~\cite{CJ1,CJ_PLB1,CJBc}.

The form factors $f_{\pm}(q^2)$ and $f_T(q^2)$ 
for $B_c(q_1\bar{q})\to P(q_2\bar{q})$ decays are obtained from the $q^+=0$ frame.
Although the form factors $f_{\pm}(q^2)$ and $f_T(q^2)$ are given in~\cite{CJBc} and~\cite{CJK},
respectively, we list them here again:
\begin{widetext}
 \bea\label{f_LFQM}
  f_{+}(q^2) &=& \int^{1}_{0}dx\int
\frac{d^{2}{\bf k}_{\perp}}{16\pi^3}
\frac{\phi_{1}(x,{\bf k}_{\perp})}{\sqrt{ {\cal A}_{1}^{2}
 + {\bf k}^{2}_{\perp}}}
\frac{\phi_{2}(x,{\bf k}'_{\perp})}{\sqrt{ {\cal A}_{2}^{2}
+ {\bf k}^{\prime 2}_{\perp}}}
( {\cal A}_{1}{\cal A}_{2}+{\bf k}_{\perp}\cdot{\bf k'}_{\perp} ),
\nonumber\\
 f_-(q^2) &=& \int^1_0 (1-x) dx
 \int \frac{ d^2{\bf k}_\perp } { 16\pi^3 }
  \frac{ \phi_1 (x, {\bf k}_\perp) } {\sqrt{ {\cal A}^2_1 + {\bf k}^2_\perp }}
  \frac{ \phi_2 (x, {\bf k'}_\perp) } {\sqrt{ {\cal A}^2_2 + {\bf k}^{\prime 2}_\perp }}
   \biggl\{ -x(1-x) M^2_1 - {\bf k}^2_\perp - m_1m + (m_2 - m){\cal A}_1
 \nonumber\\
 && + 2\frac{q\cdot P}{q^2} \biggl[ {\bf k}^2_\perp
 + 2\frac{ ( {\bf k}_\perp \cdot {\bf q}_\perp)^2 } {q^2} \biggr]
 + 2 \frac{ ( {\bf k}_\perp \cdot {\bf q}_\perp)^2 } {q^2}
  + \frac{ {\bf k}_\perp \cdot {\bf q}_\perp } {q^2}  [ M^2_2 - (1-x) (q^2 + q\cdot P) + 2 x M^2_0
  \nonumber\\
  && - (1 - 2x) M^2_1 - 2(m_1 - m) (m_1 + m_2) ] \biggr\},
  \nonumber\\
f_T(q^2) &=& (M_1 + M_2) \int^1_0 (1-x) dx
 \int \frac{ d^2{\bf k}_\perp } { 16\pi^3 }
  \frac{ \phi_1 (x, {\bf k}_\perp) } {\sqrt{ {\cal A}^2_1 + {\bf k}^2_\perp }}
  \frac{ \phi_2 (x, {\bf k'}_\perp) } {\sqrt{ {\cal A}^2_2 + {\bf k}^{\prime 2}_\perp }}
  \biggl[ (m_1 - m_2) \frac{{\bf k}_\perp\cdot{\bf q}_\perp}{ {\bf q}^2_\perp}
            + {\cal A}_1 \biggr],
  \eea
  \end{widetext}
where ${\bf k'}_\perp={\bf k}_\perp + (1-x){\bf q}_\perp$,
${\cal A}_{i}= (1-x) m_{i} + x m$ ($i=1,2$), and $q\cdot P=M^2_1-M^2_2$ with
$M_1$ and $M_2$ being the physical masses of the initial and final state mesons, respectively.
The explicit covariances of $f_{\pm}(q^2)$ and $f_T(q^2)$ are proven in~\cite{CJBc} and
in the appendix of this work, respectively.
Since the form factors $f_{\pm}(q^2)$ and $f_{T}(q^2)$ in Eq.~(\ref{f_LFQM})
are defined in the spacelike ($q^2=-{\bf q}^2_\perp <0$) region, we then analytically continue them to
the timelike $q^{2}>0$ region by changing ${\bf q}^2_{\perp}$ to $-q^2$ in
the form factors.
We should note that our analytic method 
is reliable in the entire physical region of the exclusive rare decay
since the first unitary branch point occurs just right after the zero-recoil
point, $q^2=(M_1-M_2)^2$. 

We also compare our analytic solutions with the
double pole parametric form given by
\be\label{Pole}
f_i(q^2)=\frac{f_i(0)}{1-\sigma_1 s+\sigma_2 s^2},
\ee
where $\sigma_1$ and $\sigma_2$ are the fitted parameters.

\section{Numerical results}

In our numerical calculations for the exclusive rare
$B_c\to (D,D_s)(\nu_{\ell}\bar{\nu_{\ell}})$
and $B_c\to (D,D_s)(\ell^+\ell^-)$ decays, we use
two sets of model parameters ($m_q,\beta$) for the linear and HO
confining potentials given in Table~\ref{t1}~\cite{CJ1,CJ_PLB1,CJBc}.
Although our predictions~\cite{CJBc} of ground state heavy
meson masses are overall in good agreement with the experimental
values, we use the experimental meson masses~\cite{Data08} in the
computations of the decay widths to reduce possible theoretical
uncertainties.

\begin{table}[t]
\caption{Model parameters ($m_q,\beta$) [GeV] for $D, D_s$ and $B_c$ mesons
used in our analysis. $q=u$ and $d$.}\label{t1}
\begin{tabular}{cccccccc} \hline\hline
Model & $m_q$ & $m_s$ & $m_c$ & $m_b$ & $\beta_{qc}$ & $\beta_{sc}$ & $\beta_{cb}$\\
\hline
Linear & 0.22 & 0.45 & 1.8 & 5.2  & 0.4679 & 0.5016  & 0.8068\\
\hline
HO & 0.25 & 0.48 & 1.8 & 5.2 & 0.4216 & 0.4686  & 1.0350 \\
\hline\hline
\end{tabular}
\end{table}
Note that in the numerical calculations we take
$(m_c, m_b)=(1.8,5.2)$ GeV in all formulas except in the Wilson coefficient
$C^{\rm eff}_9$, where $(m_c, m_{b,\rm pole})=(1.4,4.8)$ GeV  have been commonly
used~\cite{BM}. For the numerical values of the Wilson coefficients,
we use the results given by Ref.~\cite{BM}:
\bea\label{WC}
C_1 &=&-0.248,\; C_2=1.107, \; C_3=0.011,
\nonumber\\
C_4 &=&-0.026, \;C_5=0.007, \; C_6=-0.031,
\nonumber\\
C^{\rm eff}_7 &=&-0.313,\; C_9=4.344,\; C_{10}=-4.669,
\eea
and other input parameters are $|V_{tb}V^*_{ts}|=0.039$,
$|V_{tb}V^*_{td}|=0.008$, $\al_{\rm em}^{-1}=129$, $M_W=80.43$ GeV,
$m_t=171.3$ GeV, $\sin^2\theta_W=0.2233$,
and $\tau_{B_c}=0.46$ ps.
The effective Wilson coefficient
$C^{\rm eff}_9$ taking into account both the short distance (SD)
and the LD contributions
from $c\bar{c}$ resonance states ($J/\psi,\psi',\cdots$) has the following
form~\cite{BM}
\be
C^{\rm eff}_9(s) = C_9 + Y_{SD}(s) + Y_{LD}(s),
\ee
where the explicit forms of $Y_{SD}(s)$ and $Y_{LD}(s)$ can be found in~\cite{BM,Faessler}.
For the LD contribution $Y_{LD}(s)$, we include two $c\bar{c}$ resonant states
$J/\psi(1S)$ and $\psi'(2S)$ and use $\Gamma(J/\psi\to\ell^+\ell^-)=5.26\times 10^{-6}$
GeV, $M_{J/\psi}=3.1$ GeV, $\Gamma_{J/\psi}=87\times 10^{-6}$ GeV
for $J/\psi(1S)$ and $\Gamma(\psi'\to\ell^+\ell^-)=2.12\times 10^{-6}$
GeV, $M_{\psi'}=3.69$ GeV, $\Gamma_{\psi'}=277\times 10^{-6}$ GeV
for $\psi'(2S)$~\cite{Data08}.

\begin{figure}
\vspace{0.9cm}
\includegraphics[width=3in,height=3in]{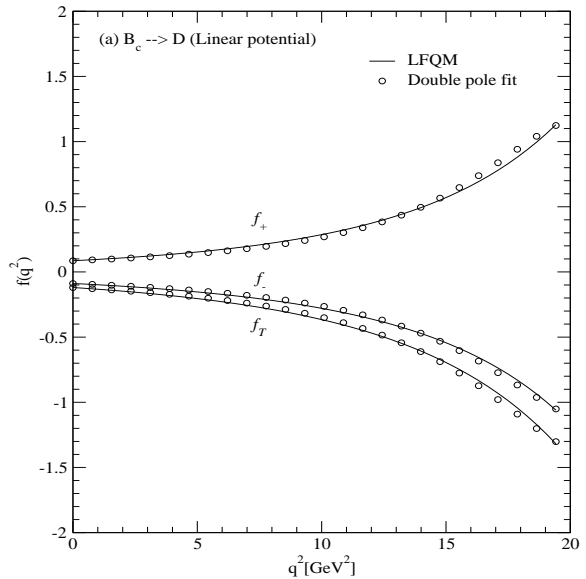}\\
\vspace{1.5cm}
\includegraphics[width=3in,height=3in]{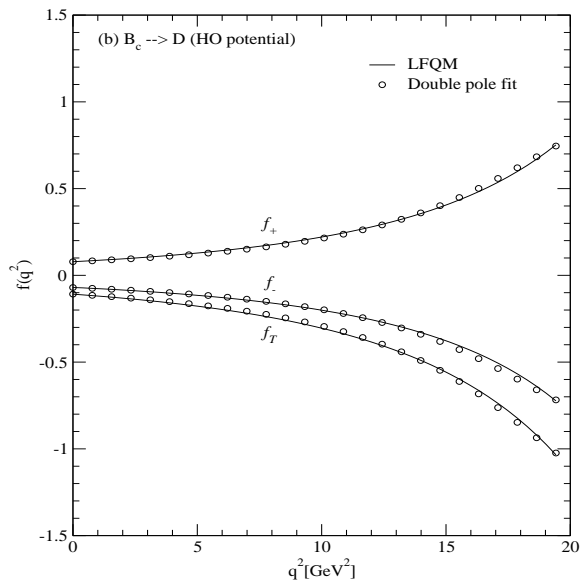}
\caption{The weak form factors (solid line)
 for $B_c\to D$ transitions obtained from the (a)
linear and (b) HO potential parameters.
The circles stand for the results from the double pole fits.}
\label{fig2}
\end{figure}
\begin{figure}
\vspace{0.9cm}
\includegraphics[width=3in,height=3in]{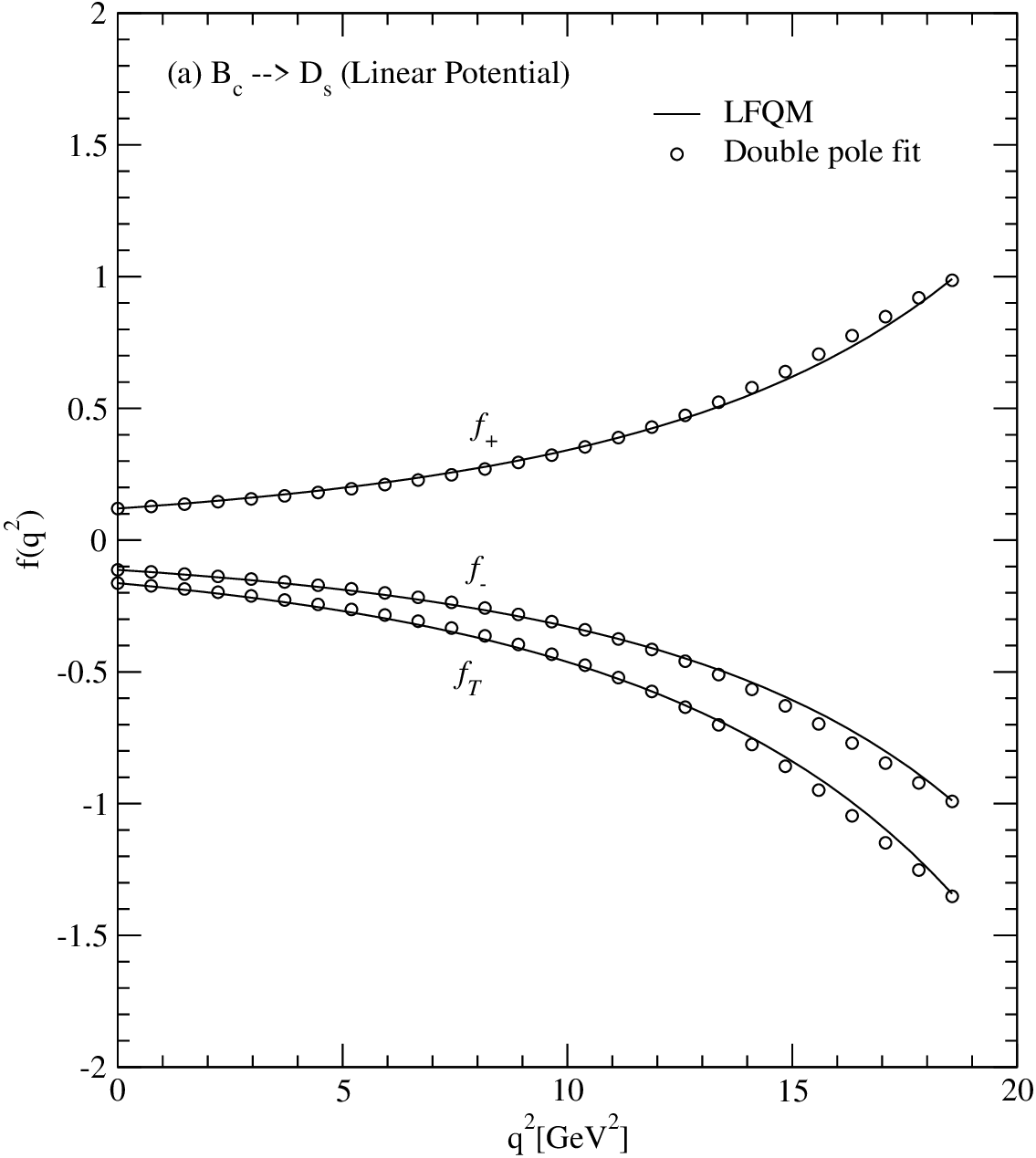}\\
\vspace{1.5cm}
\includegraphics[width=3in,height=3in]{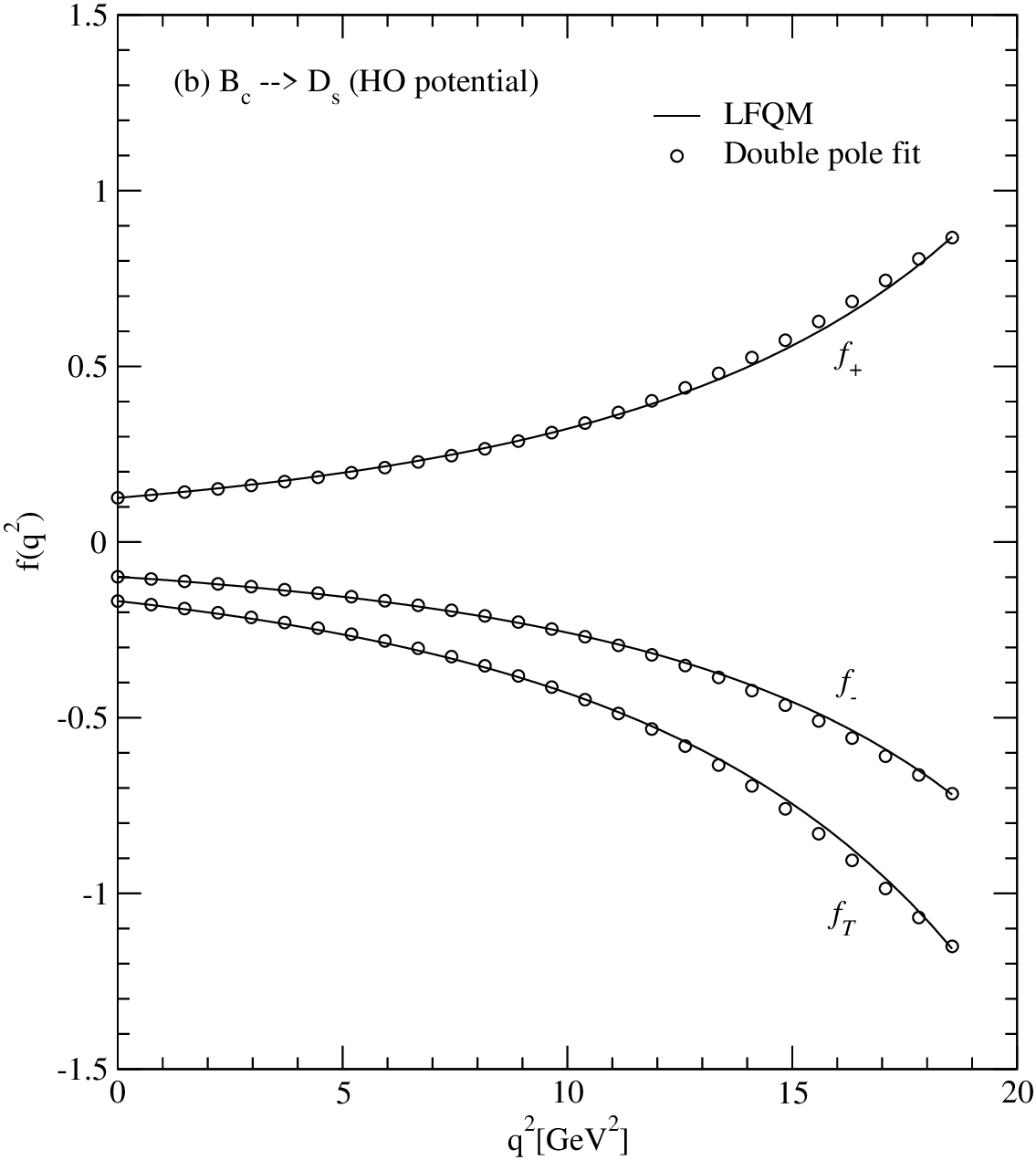}
\caption{The weak form factors (solid line)
 for $B_c\to D_s$ transitions obtained from the (a)
linear and (b) HO potential parameters.
The circles stand for the results from the double pole fits.}
\label{fig3}
\end{figure}

In Figs.~\ref{fig2} and~\ref{fig3}, we show the $q^2$ dependences of the
form factors $f_{\pm}(q^2)$ and $f_{T}(q^2)$ (solid line) for
the $B_c\to D$ and $B_c\to D_s$ transitions obtained from the (a)
linear and (b) HO potential parameters, respectively. We also include
the results (circles) obtained from the double pole form given by Eq.~(\ref{Pole}).
As one can see from Figs.~\ref{fig2} and~\ref{fig3}, our analytic solutions
are well approximated by the double pole form.

The form factors at the zero-recoil
point (i.e., $q^2=q^2_{\rm max})$ correspond to the overlap integral
of the initial and final state meson wave functions. The
maximum-recoil point (i.e., $q^2=0$) corresponds to a final state
meson recoiling with the maximum three-momentum $|{\vec
P}_{D_{(s)}}|=(M^2_{B_c}-M^2_{D_{(s)}})/2M_{B_c}$ in the rest frame of the $B_c$
meson. For $B_c\to D$ transition, while the form factors at $q^2=0$
obtained from the linear [HO] potential parameters are $f_+(0)=0.086\;[0.079]$,
$f_-(0)=-0.089\;[-0.070]$, and $f_T(0)=-0.120\;[-0.108]$,
the form factors at $q^2=q^2_{\rm max}$
are $f_+(q^2_{\rm max})=1.129 \;[0.752]$,
$f_-(q^2_{\rm max})=-1.060 \;[-0.723]$, and $f_T(q^2_{\rm max})=-1.319 \;[-1.034]$.
As for the zero-mode contribution to the form factor $f_-(q^2)$, i.e.
$f^{\rm Z.M.}_-(q^2) = f_-(q^2) - f^{\rm val}_-(q^2)$~\cite{CJBc}, we obtain the valence
contribution to $f_-(q^2)$ as
$f^{\rm val}_-(0)=-0.096\;[-0.081]$ and
$f^{\rm val}_-(q^2_{\rm max})=-1.132\;[-0.823]$ from the linear [HO] potential
parameters. This estimates about $7\%\sim 15\%$ zero-mode contribution to $f_-(q^2)$
for the $B_c\to D$ transition.
For $B_c\to D_s$ transition, while the form factors at $q^2=0$ obtained from
the linear [HO] potential parameters are $f_+(0)=0.120 \;[0.126]$,
$f_-(0)=-0.113\; [-0.099]$, and $f_T(0)=-0.163\; [-0.168]$,
the form factors at $q^2=q^2_{\rm max}$  are $f_+(q^2_{\rm max})=0.992 \;[0.868]$,
$f_-(q^2_{\rm max})=-0.988 \;[-0.719]$,
and $f_T(q^2_{\rm max})=-1.342 \;[-1.157]$. We also obtain the valence
contribution to $f_-(q^2)$ as
$f^{\rm val}_-(0)=-0.121\;[-0.105]$ and
$f^{\rm val}_-(q^2_{\rm max})=-1.114\;[-0.815]$ from the linear [HO] potential
parameters. This also estimates about $7\%\sim 15\%$ zero-mode contribution to $f_-(q^2)$
for the $B_c\to D_s$ transition.

\begin{table*}
\caption{Results for form factors at $q^2=0$ of $B_c\to D\ell^+\ell^-/\nu_{\ell}\bar{\nu_{\ell}}$
decay and parameters
$\sigma_i$ defined in Eq.~\protect(\ref{Pole}).}
\begin{tabular}{cccccccccc}
\hline\hline
Model & $f_+(0)$ & $\sigma_1$ & $\sigma_2$
      & $f_-(0)$ & $\sigma_1$ & $\sigma_2$
      & $f_T(0)$ & $\sigma_1$ & $\sigma_2$\\
\tableline
Linear & 0.086  & $-3.50$ & 3.30 & $-0.089$& $-3.38 $ & 3.09
            & $-0.120$ & $-3.35 $ & 3.06  \\
HO & 0.079 & $-3.20$ & 2.81 & $-0.070$& $-3.28$ & 2.94
            & $-0.108$ & $-3.18$ & 2.77 \\
RCQM~\protect\cite{Faessler}& 0.186 & $-3.48$ & 1.62 & $-0.190$ & $-2.44$ & 1.54
                       & $-0.275$ & $-2.40$ & 1.49\\
CQM~\protect\cite{Geng}& 0.123 & $-3.35$ & 3.03 & $-0.130$ & $-3.63$ & 3.55
                       & $-0.186$ & $-3.52$ & 3.38\\
SR~\protect\cite{Azizi}& 0.22 & $-1.10$  & $-2.48$ & $-0.29$ & $-0.63$ & $-4.06$
                       & $-0.27$ & $-0.72$ & $-3.24$\\
\hline\hline
\end{tabular}
\label{t2}
\end{table*}

\begin{table*}
\caption{Results for form factors at $q^2=0$ of $B_c\to D_s\ell^+\ell^-/\nu_{\ell}\bar{\nu_{\ell}}$
decay and parameters
$\sigma_i$ defined in Eq.~\protect(\ref{Pole}).}
\begin{tabular}{cccccccccc}
\hline\hline
Model & $f_+(0)$ & $\sigma_1$ & $\sigma_2$
      & $f_-(0)$ & $\sigma_1$ & $\sigma_2$
      & $f_T(0)$ & $\sigma_1$ & $\sigma_2$\\
\tableline
Linear & 0.120  & $-3.32$ & 3.09 & $-0.113$& $-3.36 $ & 3.14
            & $-0.163$ & $-3.28 $ & 3.00  \\
HO & 0.126 & $-3.10$ & 2.73 & $-0.099$& $-3.12$ & 2.74
            & $-0.168$ & $-3.08$ & 2.69 \\
CQM~\protect\cite{Geng}& 0.167 & $-3.40$ & 3.21 & $-0.166$ & $-3.51$ & 3.38
                       & $-0.247$ & $-3.41$ & 3.30\\
SR~\protect\cite{Azizi}& 0.16 & $-1.55$  & $-2.80$ & $-0.18$ & $-0.77$ & $-6.71$
                       & $-0.19$ & $-1.43$ & $-3.06$\\
\hline\hline
\end{tabular}
\label{t3}
\end{table*}
The form factors at $q^2=0$ and the parameters $\sigma_i$ of the double pole form
for the $B_c\to D$ and $B_c\to D_s$ transitions are listed in Tables~\ref{t2} and~\ref{t3},
respectively, and compared  with other theoretical
results~\cite{Faessler,Geng,Azizi}.
The differences of the form factors between the linear and HO potential model
predictions for the $B_c\to D$ are larger than those for the $B_c\to D_s$.
Our predictions of the form factors
are also rather smaller than other theoretical model
predictions~\cite{Faessler,Geng,Azizi}. The upcoming experimental study planned at the Tevatron
and at the LHC may distinguish these different model predictions.

\begin{figure}
\vspace{0.9cm}
\includegraphics[width=3in,height=3in]{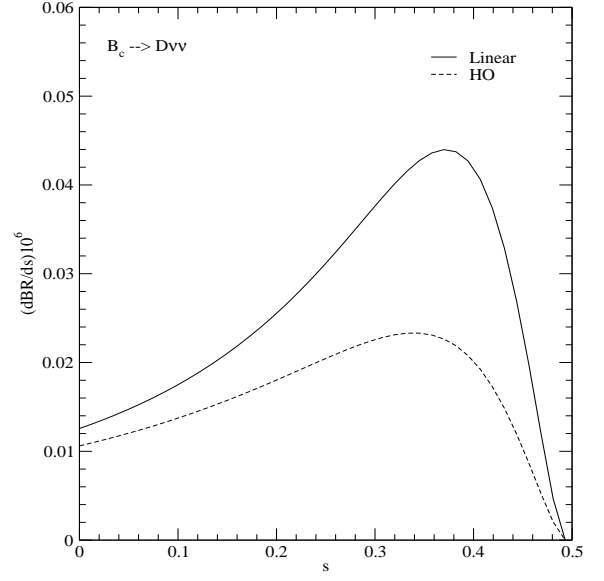}\\
\vspace{1.5cm}
\includegraphics[width=3in,height=3in]{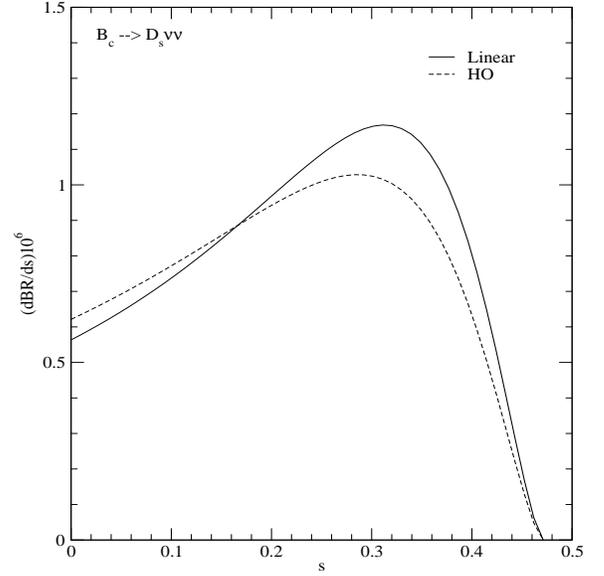}
\caption{Differential branching ratios for $B_c\to D\sum\nu_{\ell}\bar{\nu_{\ell}}$ (upper
panel) and $B_c\to D_s\sum\nu_{\ell}\bar{\nu_{\ell}}$ (lower panel) obtained from
the linear (solid line) and HO (dashed line) potential parameters.}
\label{fig4}
\end{figure}
\begin{figure}
\vspace{0.9cm}
\includegraphics[width=3in,height=3in]{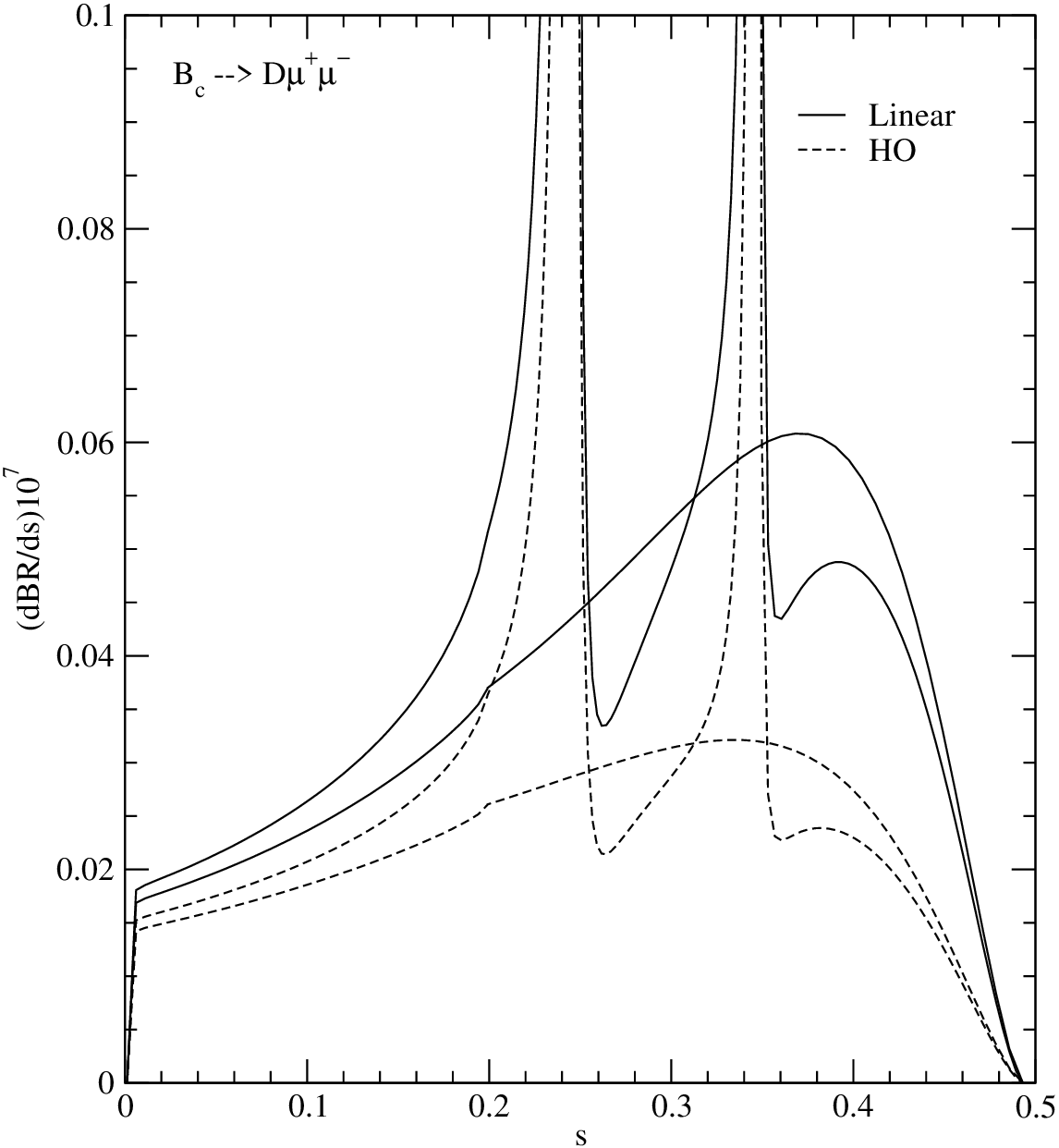}\\
\vspace{1.5cm}
\includegraphics[width=3in,height=3in]{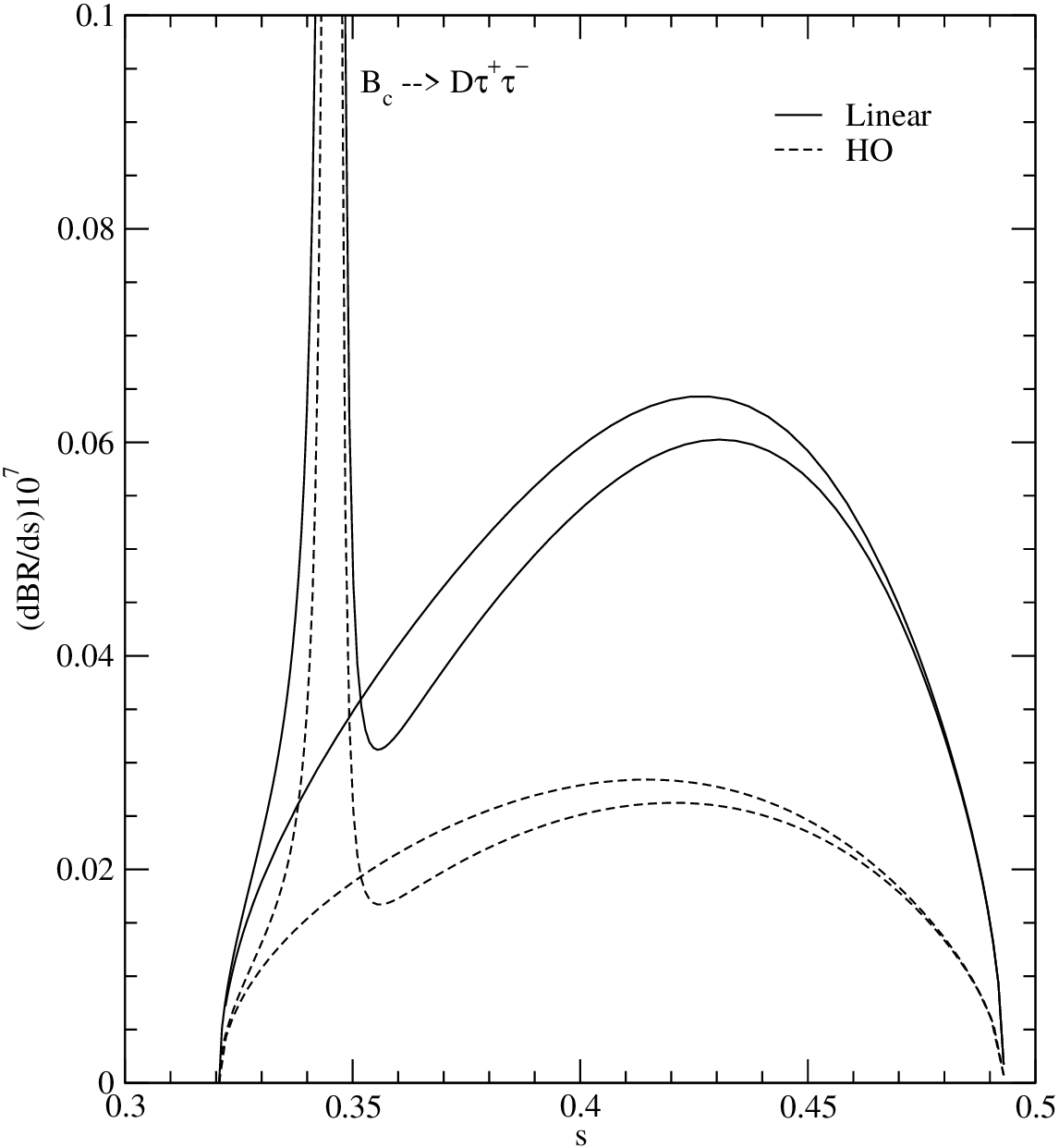}
\caption{Differential branching ratios for $B_c\to D\mu^+\mu^-$ (upper
panel) and $B_c\to D\tau^+\tau^-$ (lower panel) obtained from
the linear (solid line) and HO (dashed line) potential parameters. The
curves with (without) resonant shapes represent the results with (without)
the LD contributions.}
\label{fig5}
\end{figure}
\begin{figure}
\vspace{0.9cm}
\includegraphics[width=3in,height=3in]{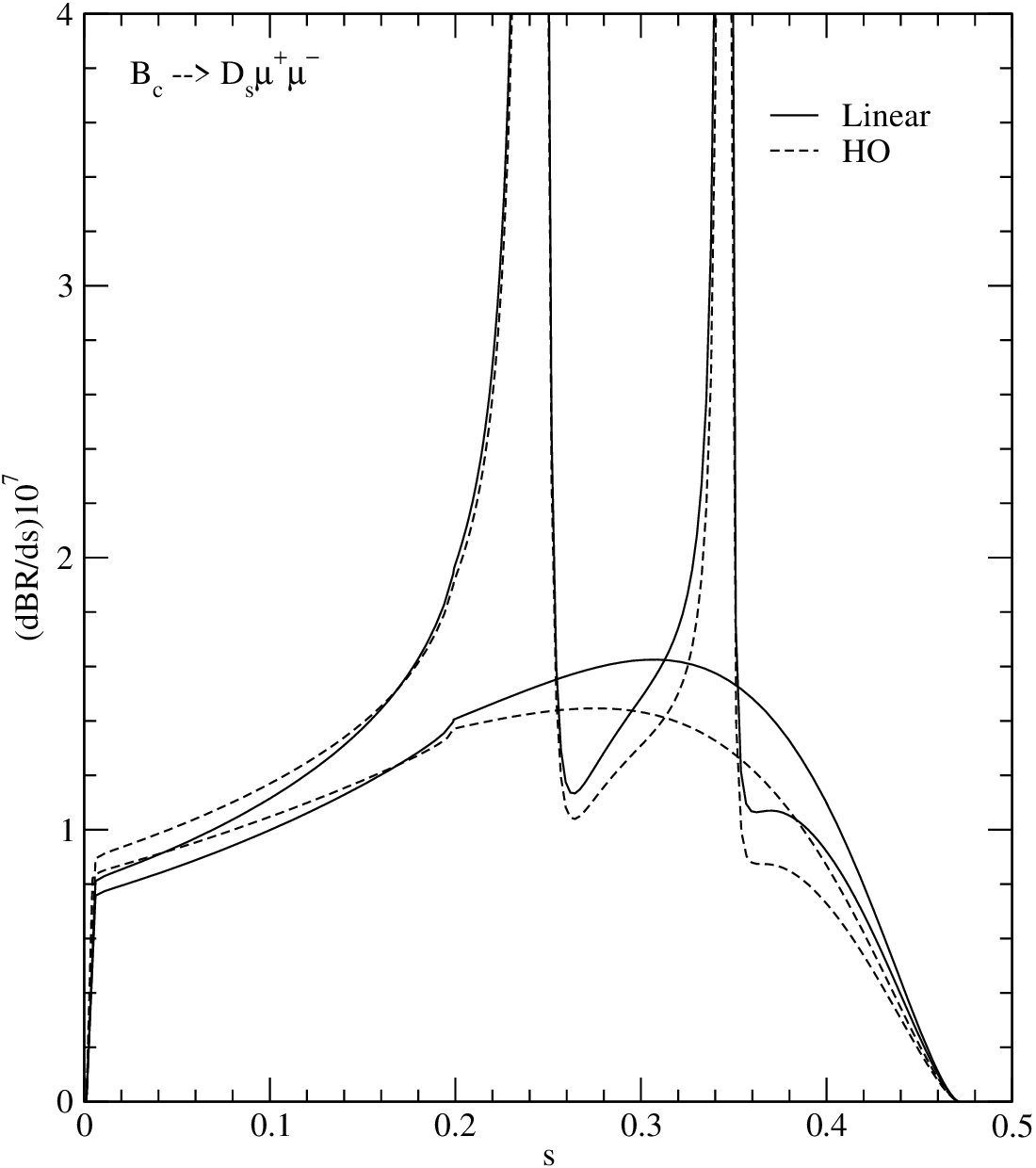}\\
\vspace{1.5cm}
\includegraphics[width=3in,height=3in]{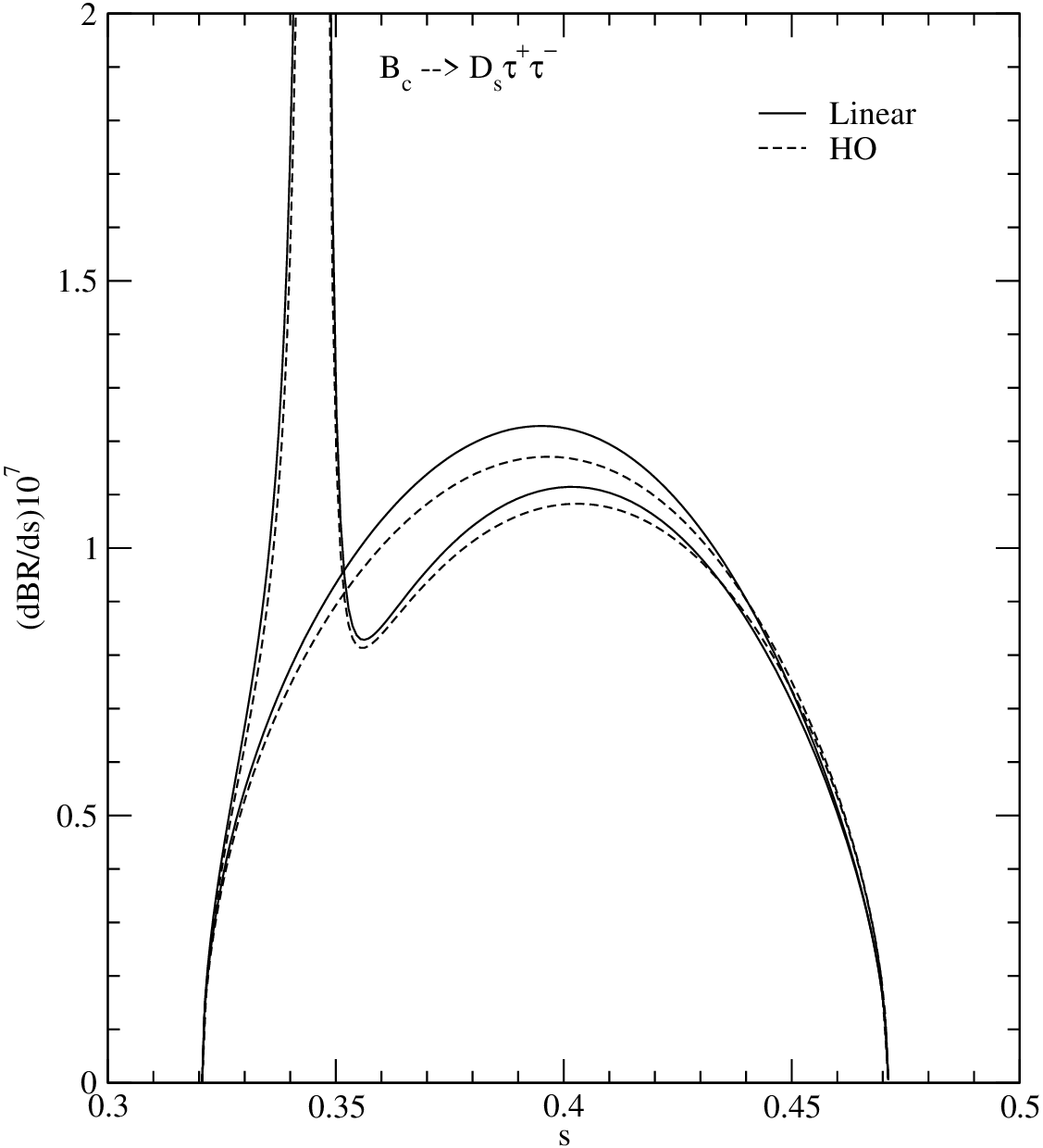}
\caption{Differential branching ratios for $B_c\to D_s\mu^+\mu^-$ (upper
panel) and $B_c\to D_s\tau^+\tau^-$ (lower panel) obtained from
the linear (solid line) and HO (dashed line) potential parameters.}
\label{fig6}
\end{figure}

We show our results for the differential branching ratios for
$B_c\to (D, D_s)\sum\nu_{\ell}\bar{\nu}_{\ell}$ in Fig.~\ref{fig4},
$B_c\to D \ell^+\ell^-$ in Fig.~\ref{fig5}, and $B_c\to D_s\ell^+\ell^-$
in Fig.~\ref{fig6}, respectively.
The solid (dashed) line represents the
result obtained from the linear (HO) potential parameters. For the
$B_c\to (D, D_s)\ell^+\ell^-$ transitions in Figs.~\ref{fig5} and~\ref{fig6},
the curves with (without) resonant
shapes represent the results with (without) the LD contribution to $C^{\rm eff}_9$.
Although the form factor $f_-(q^2)$ does not contribute to the branching
ratio in the massless lepton ($\ell=e$ or $\mu$) decay, it is necessary for
the heavy $\tau$ decay process.
As one can see from Figs.~\ref{fig5} and~\ref{fig6},
the LD contributions clearly overwhelm the branching
ratios near $J/\psi(1S)$ and $\psi'(2S)$ peaks, however, suitable
$\ell^+\ell^-$ invariant mass cuts can separate the LD contribution
from the SD one away from these peaks.
This divides the spectrum into two distinct regions~\cite{Hew,AGM}:
(i) low-dilepton mass, $4m^2_\ell\leq q^2\leq M^2_{J/\psi}-\delta$,
and (ii) high-dilepton mass,
$M^2_{\psi'}+\delta\leq q^2\leq q^2_{\rm max}$, where $\delta$ is to
be matched to an experimental cut.

Our predictions for the nonresonant branching ratios
obtained from the linear and the HO potential parameters are summarized in Table~\ref{t4}
and compared with other theoretical predictions~\cite{Faessler,Geng,Azizi} within
the SM. Since the amplitude $B_c\to (D, D_s)\ell^+\ell^-$ is regular at $q^2=0$, the
transitions $B_c\to (D, D_s)e^+e^-$ and $B_c\to (D, D_s)\mu^+\mu^-$ have almost
the same decay rates, i.e. insensitive to the mass of the light lepton.
The branching ratios with the LD contributions
for $B_c\to (D,D_s)\ell^+\ell^-$ $(\ell=\mu,\tau)$ obtained from the
linear (HO) potential parameters are also presented in Table~\ref{t5}
for low- and high-dilepton mass regions of $q^2$.
\begin{table}
\caption{Nonresonant branching ratios (in units of $10^{-8}$)
for $B_c\to (D, D_s)(\nu_{\ell}\bar{\nu}_{\ell})$ and
$B_c\to (D, D_s)(\ell^+\ell^-)$ transitions compared with other theoretical model
predictions within the SM.}
\begin{tabular}{lccccc}
\hline\hline
Mode & Linear & HO & ~\protect\cite{Faessler} & ~\protect\cite{Geng}
& ~\protect\cite{Azizi} \\
\hline
$B_c\to D\sum\nu_{\ell}\bar{\nu}_{\ell}$ & 1.31 &  0.81 & 3.28 & 2.74 & $(3.38\pm 0.71)$ \\
$B_c\to D_s\sum\nu_{\ell}\bar{\nu}_{\ell}$ & 39 &  37 & 73 & 92 & $(49\pm 12)$ \\
$B_c\to D\mu^+\mu^-$ & 0.18 &  0.11 & 0.44 & 0.40 & $(0.31\pm 0.06)$ \\
$B_c\to D_s\mu^+\mu^-$ & 5.4 &  5.1 & 9.7 & 13.3 & $(6.1\pm 1.5)$ \\
$B_c\to D\tau^+\tau^-$ & 0.08 &  0.04 & 0.11 & 0.12 & $(0.13\pm 0.03)$ \\
$B_c\to D_s\tau^+\tau^-$ & 1.4 &  1.3 & 2.2 & 3.7 & $(2.3\pm 0.5)$ \\
\hline\hline
\end{tabular}
\label{t4}
\end{table}
\begin{table}
\caption{Branching ratios with the LD contributions
for $B_c\to (D,D_s)\ell^+\ell^-$ for low and high dilepton mass
regions of $q^2$ [GeV$^2$] obtained from the
linear (HO) potential parameters.}
\begin{tabular}{ccc}
\hline\hline
Mode & $4m^2_\ell\leq q^2\leq 8.5$&   $13.8\leq q^2\leq q^2_{\rm max}$ \\
\hline
$B_c\to D\mu^+\mu^-$  & $6.56 \;(4.98)\times10^{-10}$ & $5.00 \;(2.38)\times10^{-10}$ \\
$B_c\to D\tau^+\tau^-$  &  & $6.68 \;(3.01)\times10^{-10}$ \\
$B_c\to D_s\mu^+\mu^-$  & $2.68 \;(2.75)\times10^{-8}$ & $0.86 \;(0.69)\times10^{-8}$ \\
$B_c\to D_s\tau^+\tau^-$  &  & $1.09 \;(1.07)\times10^{-8}$ \\
\hline\hline
\end{tabular}
\label{t5}
\end{table}

\begin{figure}
\vspace{0.9cm}
\includegraphics[width=3in,height=3in]{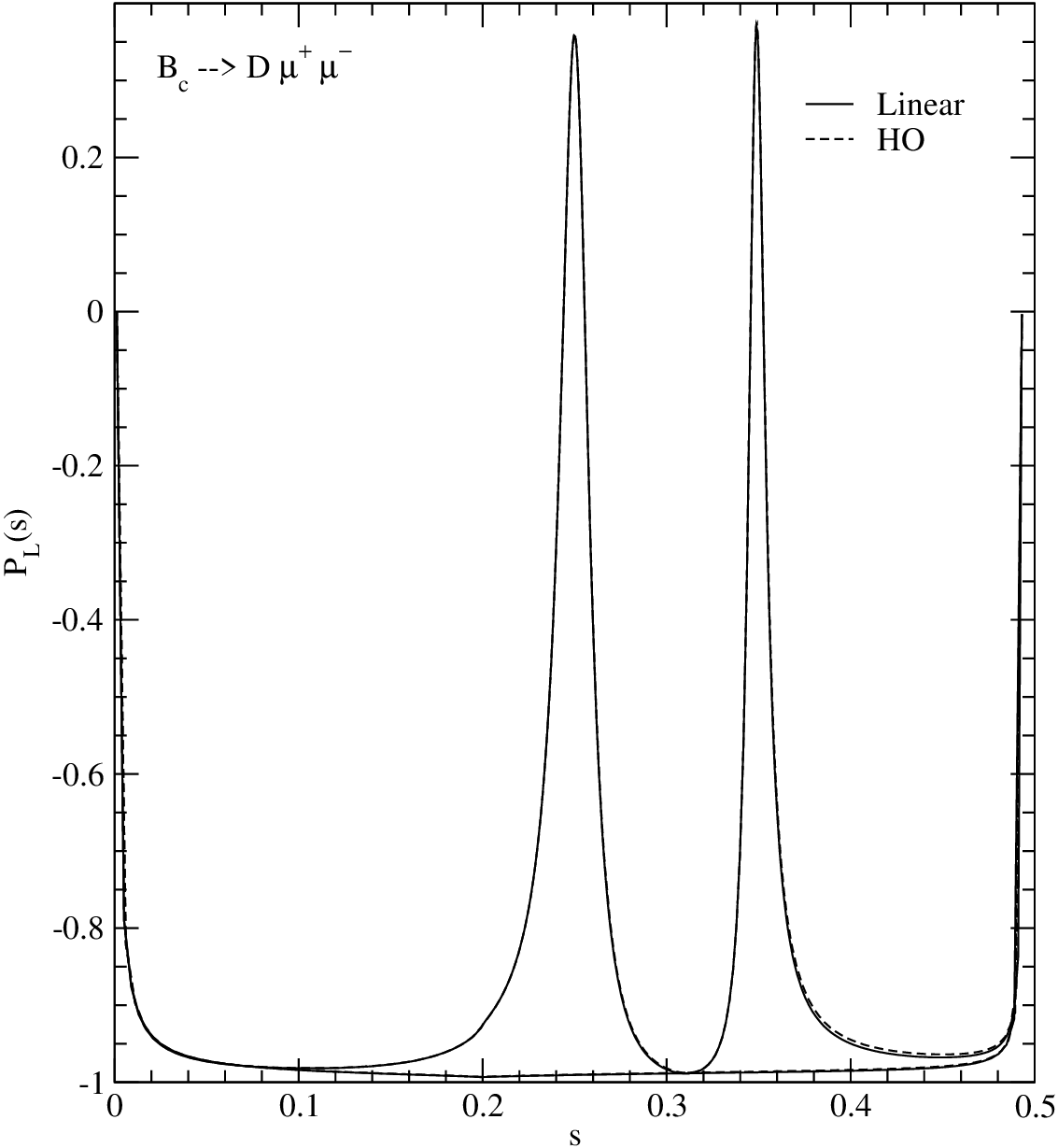}\\
\vspace{1.5cm}
\includegraphics[width=3in,height=3in]{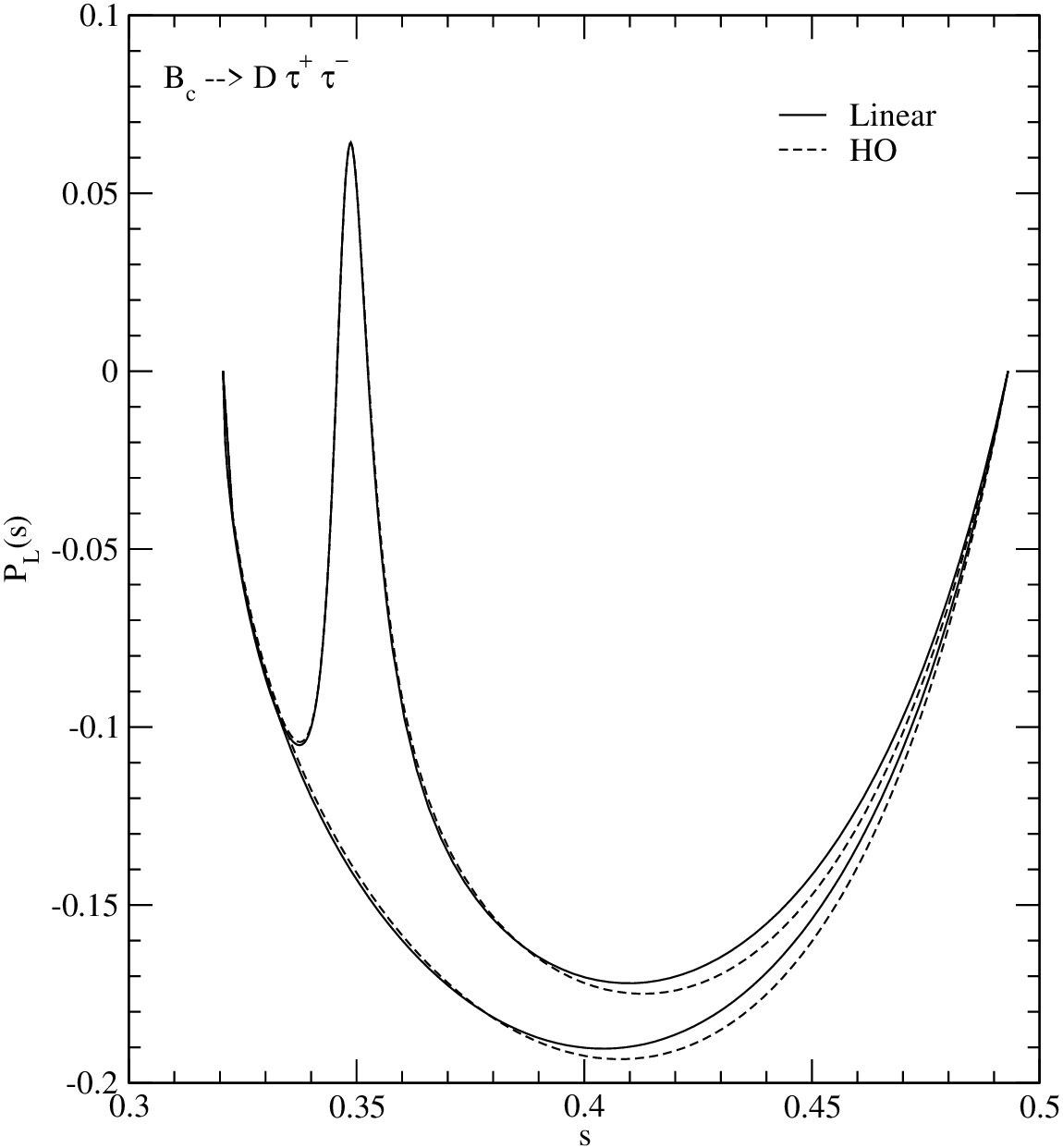}
\caption{Longitudinal lepton polarization asymmetries for $B_c\to D\mu^+\mu^-$(upper
panel) and $B_c\to D\tau^+\tau^-$(lower panel) obtained from
the linear (solid line) and HO (dashed line) potential parameters.}
\label{fig7}
\end{figure}

\begin{figure}
\vspace{0.9cm}
\includegraphics[width=3in,height=3in]{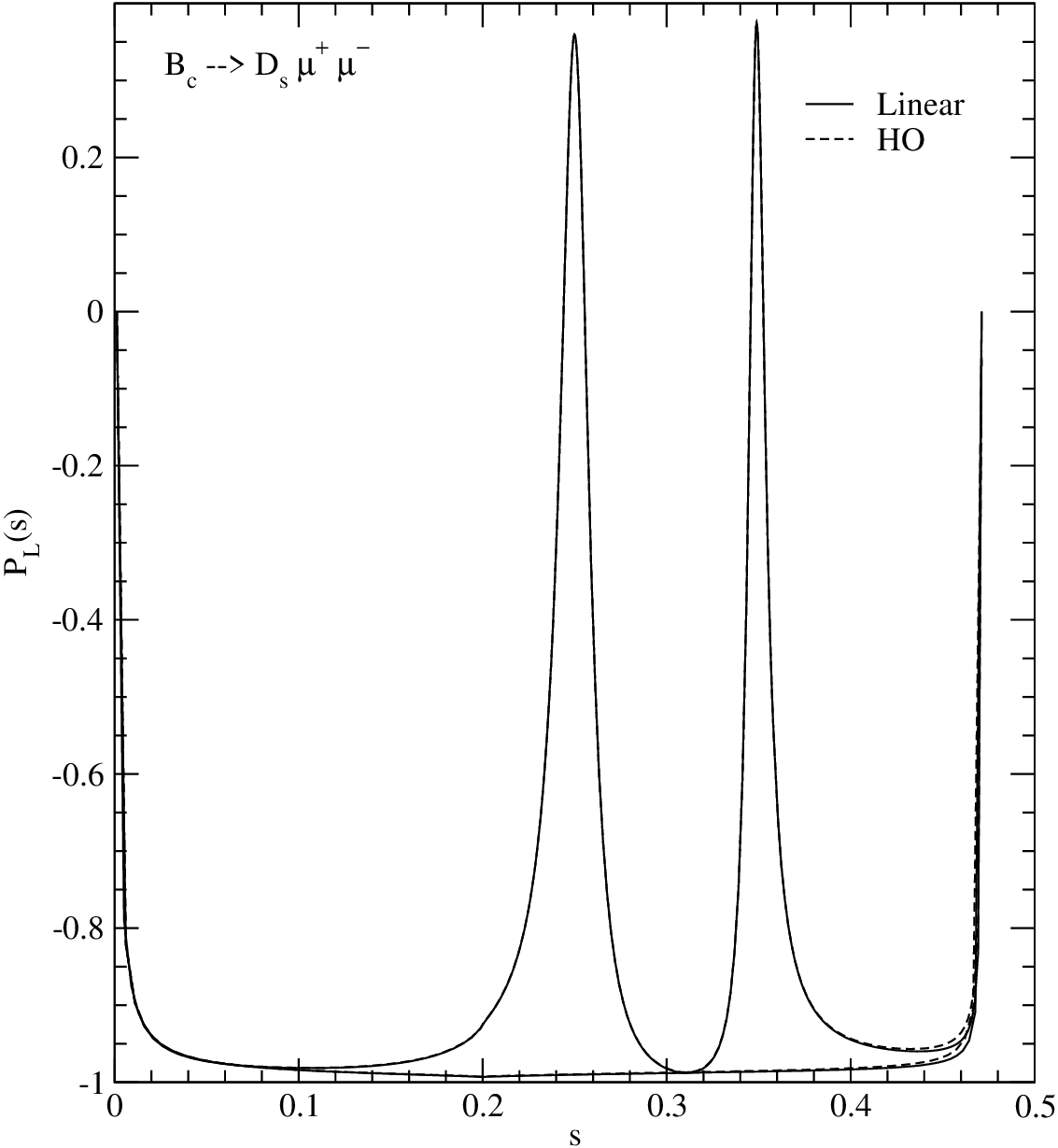}\\
\vspace{1.5cm}
\includegraphics[width=3in,height=3in]{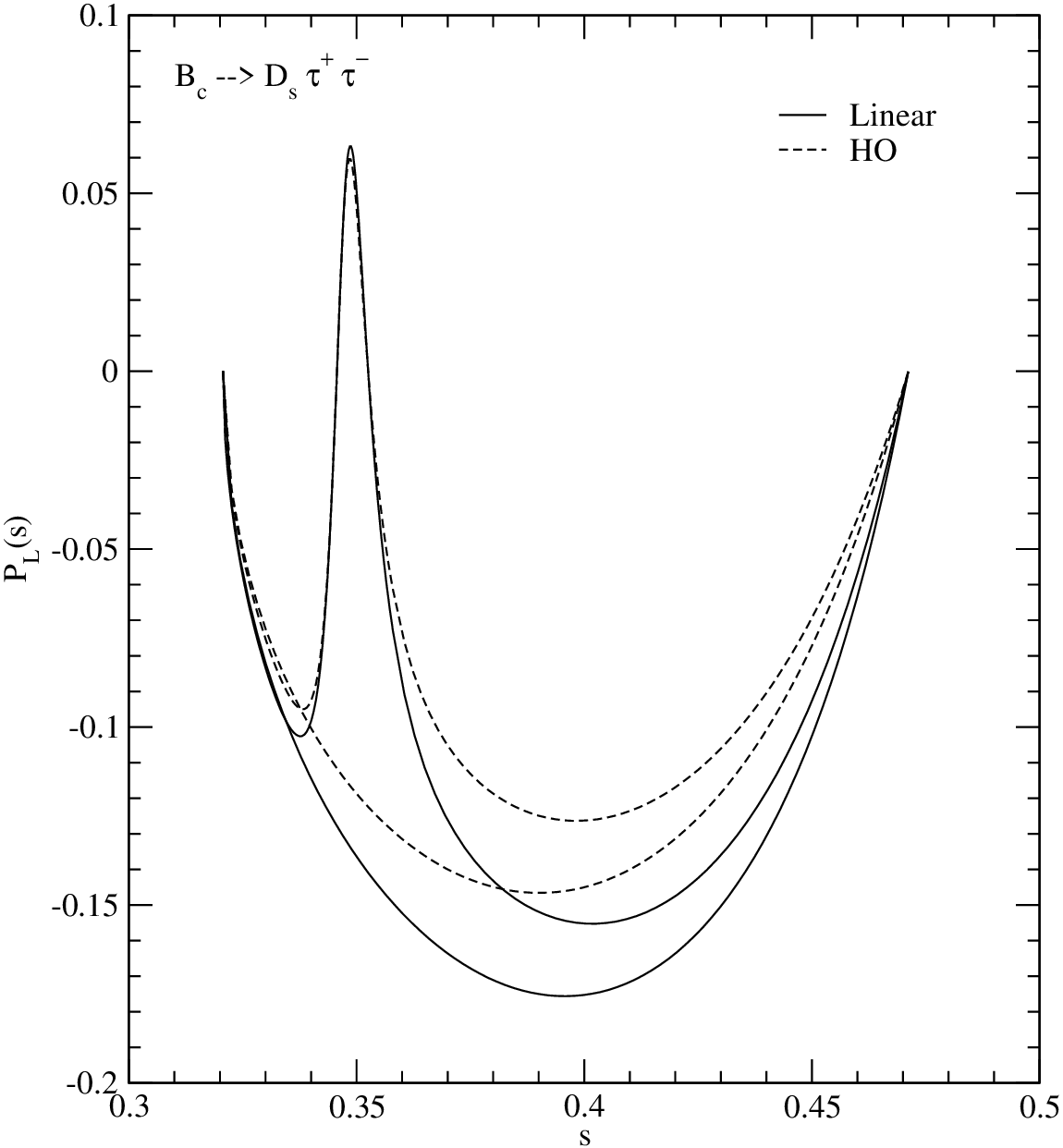}
\caption{Longitudinal lepton polarization asymmetries for $B_c\to D_s\mu^+\mu^-$(upper
panel) and $B_c\to D_s\tau^+\tau^-$(lower panel) obtained from
the linear (solid line) and HO (dashed line) potential parameters.}
\label{fig8}
\end{figure}

In Figs.~\ref{fig7} and~\ref{fig8}, we show the
LPAs  for $B\to D\ell^+\ell^-$
and $B\to D_s\ell^+\ell^-$ ($\ell=\mu,\tau$) as a function of $s$, respectively, obtained
from the linear (solid line) and HO (dashed line) potential parameters.
The curves with (without) resonant shapes represent the results with (without)
the LD contributions. In both figures, the LPAs
become zero at the end point regions of $s$. However, we note that if $m_\ell =0$, the LPAs
are not zero at the end points. As in the case of the $B\to K\mu^+\mu^-$ decay where $P_L\simeq -1$
away from the end point regions~\cite{CJK,GK,MN,HQ}, the LPAs
away from the end point regions are close to $-1$ for both $B_c\to D\mu^+\mu^-$
and $B_c\to D_s\mu^+\mu^-$ transitions.
In fact, the $P_L$ for the muon decay is insensitive to the form factors,
e.g. our $P_L$ away from the end point regions is well
approximated by~\cite{HQ}
\be\label{PLmu}
P_L\simeq 2\frac{C_{10}{\rm Re}C^{\rm eff}_9}{|C^{\rm eff}_9|^2
+ |C_{10}|^2}\simeq -1,
\ee
in the limit of $C^{\rm eff}_7\to 0$ from Eq.~(\ref{LPA_Bc}). It also shows that
the LPA for the $\mu$ dilepton channel
is insensitive to the little variation of $C^{\rm eff}_7$ as expected.
On the other hand, the LPA for the $\tau$ dilepton channel is somewhat sensitive
to the form factors.

The averaged values of $P_L$ without the LD contributions
obtained from the linear (HO) potential parameters are
$\la P_L(B_c\to D\mu^+\mu^-)\ra=-0.99\;(-0.99)$,
$\la P_L(B_c\to D\tau^+\tau^-)=-0.16\;(-0.15)$,
$\la P_L(B_c\to D_s\mu^+\mu^-)\ra=-0.98\;(-0.98)$
and $\la P_L(B_c\to D_s\tau^+\tau^-)=-0.14\;(-0.12)$, respectively.


\section{Summary and Discussion}
In this work, we investigated the exclusive rare semileptonic
$B_c\to (D,D_s)\nu_{\ell}\bar{\nu_{\ell}}$ and $B_c\to
(D,D_s)\ell^+\ell^-$ ($\ell=e,\mu,\tau$) decays within the SM,
using  our LFQM constrained by the variational principle for the QCD
motivated effective Hamiltonian with the linear ( or HO) plus Coulomb interaction.
Our model parameters obtained from the
variational principle uniquely determine the physical quantities
related to the above processes. This approach can establish the
broader applicability of our LFQM to the wider range of hadronic
phenomena. For instance, our LFQM has been tested extensively in
the spacelike processes~\cite{CJ1} as well as in the
timelike exclusive processes such as
semileptonic~\cite{CJ_PLB1,JC_E,CJBc}, rare semileptonic~\cite{CJK},
radiative~\cite{ChoiRD},
and nonleptonic two-body~\cite{CJNRD} decays of
pseudoscalar and vector mesons.

The weak form factors $f_{\pm}(q^2)$ and $f_T(q^2)$ for the rare semileptonic decays
between two pseudoscalar mesons  are
obtained in the $q^+=0$ frame ($q^2=-{\bf q}^2_\perp<0$) and then
analytically continued to the timelike region by changing ${\bf
q}^2_\perp$ to $-q^2$ in the form factor. The covariance (i.e.,
frame independence) of our model has been checked by performing the
LF calculation in the $q^+=0$ frame in parallel with the manifestly
covariant calculation using the exactly solvable covariant fermion
field theory model in $(3+1)$-dimensions.
While the form factors $f_+(q^2)$ and $f_T(q^2)$ are immune to the zero modes,
the form factor $f_-(q^2)$ is not free from the zero mode.
Our numerical results show that the zero-mode contribution to the form factor
$f_-(q^2)$ amounts to $7\%\sim 15\%$ for both $B_c\to D$ and $B_c\to D_s$
decays.

Using the solutions of the weak form factors obtained from the $q^+=0$ frame,
we calculated the branching ratios for
$B_c\to (D,D_s)\nu_{\ell}\bar{\nu_{\ell}}$ and $B_c\to
(D,D_s)\ell^+\ell^-$ and the LPAs
for $B_c\to (D,D_s)\ell^+\ell^-$ including both
short- and long-distance contributions from the QCD Wilson coefficients.
Our numerical results for the nonresonant branching ratios for
$B_c\to D(D_s)\sum\nu_{\ell}\bar{\nu_{\ell}}$ and
$B_c\to D(D_s)\ell^+\ell^-$
are in the order of $10^{-8}\;(10^{-7})$ and $10^{-9}\;(10^{-8})$, respectively.
The averaged values of the  LPAs
obtained from the linear (HO) potential parameters
are found to be $-0.99\;(-0.99)$ for $B_c\to D\mu^+\mu^-$  and $-0.16\;(-0.15)$ for
$B_c\to D\tau^+\tau^-$, and $-0.98\;(-0.98)$ for $B_c\to D_s\mu^+\mu^-$ and $-0.14\;(-0.12)$ for
$B_c\to D_s\tau^+\tau^-$, respectively. These polarization asymmetries provide valuable
information on the flavor changing loop effects in the SM.
Although the $q^2$ dependent behaviors of our form factors for the $B_c\to D_{(d,s)}$
transitions are not much different
from those of other theoretical predictions~\cite{Faessler,Geng,Azizi}, our results
for the decay rates are slightly less than those of Refs.~\cite{Faessler,Geng,Azizi}.
This difference essentially comes from the different values of the form factors
at the maximum recoil point and may be tested by future experiments.
The decay rates for the $B_c\to D\ell^+\ell^-$ and the LPAs for the
$B_c\to D_s\tau^+\tau^-$
are also quite sensitive to the choice of potential within our LFQM.
From the future experimental data on these sensitive processes, one may obtain
more realistic information on the potential between quark and antiquark in the
heavy meson system.

\acknowledgments This work  was supported by the Korea
Research Foundation Grant funded by the Korean
Government(KRF-2008-521-C00077).

\appendix*
\section{LF covariance of tensor form factor $f_T(q^2)$}

In the solvable model, based on the covariant Bethe-Salpeter (BS) model of
($3+1$)-dimensional fermion field theory~\cite{BCJ01,MF,Jaus99}, the matrix element
$J^\mu_{T}$ of the tensor current (see Eq.~(\ref{JTmu}))
is given by
 \be\label{ap:3}
 J^\mu_{T} = ig_1 g_2
\Lambda^2_1\Lambda^2_2\int\frac{d^4k}{(2\pi)^4} \frac{S^\mu_{T}} {N_{\Lambda_1}
N_{1} N_{k} N_{2} N_{\Lambda_2}},
 \ee
where $g_1$ and $g_2$ are the normalization factors which can be fixed by
requiring both charge form factors of pseudoscalar mesons to be unity at zero
momentum transfer, respectively. The denominators in Eq.~(\ref{ap:3}),
are given by
\bea\label{ap:4}
N_{k} &=& k^2 - m^2 + i\ep, \; N_{j} = p_j^2 -{m_j}^2 + i\ep,
\nonumber\\
N_{\Lambda_j} &=& p_j^2-{\Lambda_j}^2+i\ep \;(j=1,2),
 \eea
where $m_1$, $m$, and  $m_2$ are the masses of the constituents carrying the intermediate
four-momenta $p_1=P_1 -k$, $k$, and $p_2=P_2 -k$, respectively.
$\Lambda_1$ and $\Lambda_2$ play the role of momentum cut-offs similar to the Pauli-Villars
regularization~\cite{BCJ01}. The trace term $S^\mu_T$ is given by
\bea\label{ap:5T}
 S^\mu_T &=& {\rm Tr}[\gamma_5(\not\!p_1 + m_1)i\sigma^{\mu\nu}q_\nu (\not\!p_2
+m_2)\gamma_5(-\not\!k + m)]
 \nonumber\\
 &=&-4 \{ p^\mu_1 [m(p_2\cdot q) + m_2(k\cdot q)] - p^\mu_2[ m(p_1\cdot q) \nonumber\\
 &&  + m_1(k\cdot q)]
 + k^\mu [m_1(p_2\cdot q) - m_2(p_1\cdot q)] \}.
 \eea
Following the same procedure as in~\cite{CJBc}
for the calculation of the form factors $f_\pm(q^2)$, we
obtain the manifestly covariant form factor $f_T(q^2)$ as follows:

\bea\label{ap:8}
f^{\rm Cov}_T(q^2) &=& \frac{N (M_1+M_2)}{8\pi^2(\Lambda^2_1-m^2_1)(\Lambda^2_2-m^2_2)}
\int^1_0 dx\int^{1-x}_0 dy
\nonumber\\
&&\times
  [xm_1 + ym_2 + (1-x-y)m ]C,
\eea
where $N= g_1 g_2\Lambda^2_1\Lambda^2_2$ and $C$ is given by Eq.~(14) of Ref.~\cite{CJBc}.

Performing the LF calculation of Eq.~(\ref{ap:3}) in the $q^+=0$ frame, 
we use the plus component of the currents
to obtain the form factor $f_T(q^2)$, i.e.,
 \bea\label{ap:17}
  f^{\rm LF}_T(q^2) = (M_1 + M_2)\frac{J^+_T}{2q^2P^+_1}.
 \eea
The LF calculation for $S^\mu_T$ in~(\ref{ap:5T}) can be separated into the
on-mass shell ( $p^- = p^-_{\rm on}$)
propagating part $S^{\mu}_{\rm on}$ and the (off-mass shell) instantaneous one
$S^{\mu}_{\rm inst}$ using the following identity:
 \be\label{ap:10}
 \not\!p + m = (\not\!p_{\rm on} + m) + \frac{1}{2}\gamma^+(p^- - p^-_{\rm on}).
 \ee
Then the trace term $S^\mu_T$ in Eq.~(\ref{ap:5T})
is given by
 \be\label{ap:11}
   S^\mu_{T}  = S^\mu_{T\rm on} + S^\mu_{T\rm inst},
 \ee
where the on-mass shell propagating part $S^\mu_{T\rm on}$ has the same form as
$S^\mu_T$ in Eq.~(\ref{ap:5T}) but with $p^- = p^-_{\rm on}$.
The instantaneous part $S^{\mu}_{T\rm inst}$ is given by
 \bea\label{ap:13b}
  S^\mu_{T\rm inst} &=&
  -2g^{\mu +} \Delta p^-_1 [m (p_{2\rm on}\cdot q) + m_2 (k_{\rm on}\cdot q) ]
  \nonumber\\
   &+& 2g^{\mu +} \Delta p^-_2 [m (p_{1\rm on}\cdot q) + m_1 (k_{\rm on}\cdot q) ]
  \nonumber\\
  &+& 2g^{\mu +} \Delta k^-  [m_2 (p_{1\rm on}\cdot q) - m_1 (p_{2\rm on}\cdot q) ],
  \nonumber\\
 \eea
where $\Delta p^-_i= p^-_i - p^-_{i\rm on}$.
By doing the integration over $k^-$ in Eq.~(\ref{ap:3}), one finds the two LF time-ordered
contributions to the residue calculations corresponding to the two poles in $k^-$,
the LF valence contribution defined in $0<k^+<P^+_2$ region and the
nonvalence contribution  defined in $P^+_2<k^+<P^+_1$ region.
The nonvalence contribution 
in the $q^+>0$ frame corresponds to the zero mode (if it exists) in the $q^+\to 0$ limit.
As we have shown in~\cite{CJBc,BCJ01}, the LF valence contribution
comes exclusively from the on-mass shell propagating part and the zero mode
from the instantaneous part. This implies that the form factor $f_T(q^2)$ is 
free from the zero mode since $S^+_{T\rm inst}=0$.
The  LF form factor $f^{\rm LF}_T(q^2)$ is then obtained as 
\bea\label{ap:29}
  f^{\rm LF}_T(q^2)&=& \frac{N}{8\pi^3}(M_1 + M_2)\int^1_0 \frac{dx}{(1-x)}
 \int d^2{\bf k}_\perp \chi_1 (x, {\bf k}_\perp)
 \nonumber\\
 &&\times
  \chi_2(x, {\bf k'}_\perp)
   \biggl[ (m_1 - m_2) \frac{{\bf k}_\perp\cdot{\bf q}_\perp}{ {\bf q}^2_\perp}
            + {\cal A}_1 \biggr],
 \eea
 where ${\bf k'}_\perp={\bf k}_\perp + (1-x) {\bf q}_\perp$,
 ${\cal A}_i=(1-x)m_i + xm (i=1,2)$, and $q\cdot P = M^2_1 - M^2_2$.
 The LF vertex functions $\chi_1$ and $\chi_2$ are given by
 \bea\label{ap:21}
 \chi_1(x,{\bf k}_\perp) &=& \frac{1}{ x^2 (M^2_1 - M^2_0)(M^2_1-M^2_{\Lambda_1})},
 \nonumber\\
\chi_2 (x, {\bf k'}_\perp) &=& \frac{1}{ x^2 (M^2_2 -
M'^2_0)(M^2_2-M'^2_{\Lambda_2})},
\eea
where
 \bea\label{ap:19}
M^2_0 &=& \frac{{\bf k}^2_\perp + m^2}{1-x} + \frac{{\bf k}^2_\perp + m^2_1}{x},
\nonumber\\
M'^2_0 &=& \frac{{\bf k'}^2_\perp + m^2}{1-x} + \frac{{\bf k'}^2_\perp + m^2_2}{x},
 \eea
and $M^2_{\Lambda_1}=M^2_0(m_1\to\Lambda_1)$, $M'^2_{\Lambda_2}=
M'^2_0(m_2\to\Lambda_2)$. We numerically confirm that our LF form factor
$f^{\rm LF}_T(q^2)$ is exactly the same as the manifestly 
covariant form factor $f^{\rm Cov}_T(q^2)$. This proves that $f^{\rm LF}_T(q^2)$ is
immune to the zero mode.

Following the same procedure~\cite{CJBc} to obtain the form factors $f_\pm(q^2)$ within our LFQM,
the form factor $f_T(q^2)$ given by Eq.~(\ref{f_LFQM}) is obtained by the following relations:
\bea\label{ap:33}
 \sqrt{2N} \frac{ \chi_1(x,{\bf k}_\perp) } {1-x}
 &=& \frac{ \phi_1(x,{\bf k}_\perp) } {\sqrt{ {\cal A}^2_1 + {\bf k}^2_\perp }},
 \nonumber\\
\sqrt{2N} \frac{ \chi_2(x,{\bf k'}_\perp) } {1-x}
 &=& \frac{ \phi_2(x,{\bf k'}_\perp) } {\sqrt{ {\cal A}^2_2 + {\bf k'}^2_\perp }}.
 \eea

\end{document}